%% file: 00-main.tex
\documentclass[conference]{IEEEtran}
\input{01-packages.tex}

\begin{document}
\input{03-title-authors.tex}

\maketitle


\input{05-abstract.tex}
\input{10-intro.tex}

\input{20-methodology.tex}

\input{30-experiment.tex}

\input{40-conclusion.tex}

\input{90-acknowledgements.tex}

\small
\bibliographystyle{IEEEtran}
\bibliography{99-references}

\end{document}

%% file: 01-packages.tex
\usepackage{cite}
\usepackage{graphicx}

\usepackage{todonotes}
\usepackage{blkarray}
\usepackage{hyperref, url} 
\usepackage{xcolor}

\usepackage{amsmath, amssymb, amsfonts, amscd} 

\usepackage{lipsum} 

\usepackage{courier} 

\usepackage{wrapfig}
\usepackage{threeparttable} 
\usepackage{rotating} 
\usepackage{subcaption}
\usepackage{float}
\graphicspath{ {./images/} } 
\usepackage{multicol}
\usepackage{colortbl} 
\definecolor{lightred}{rgb}{ .98, .79,  .79} 
\definecolor{darkolivegreen}{rgb}{0.33, 0.42, 0.18} 

\usepackage{color} 

\usepackage[numbers,sort&compress,square]{natbib} 

\usepackage{enumitem} 

\IEEEoverridecommandlockouts 

%% file: 03-title-authors.tex
\title{\LARGE{Exploiting the Shape of CAN Data for In-Vehicle Intrusion Detection}\thanks{
\scriptsize{This manuscript has been authored by UT-Battelle, LLC under Contract No. DE-AC05-00OR22725 with the US DOE. The United States Government retains and the publisher, by accepting the article for publication, acknowledges that the United States Government retains a non-exclusive, paid-up, irrevocable, world-wide license to publish or reproduce the published form of this manuscript, or allow others to do so, for United States Government purposes. The DOE will provide public access to these results of federally sponsored research in accordance with the DOE Public Access Plan (\url{http://energy.gov/downloads/doe-public-access-plan).} }
}
}
\author{
\IEEEauthorblockN{
Zachariah Tyree\IEEEauthorrefmark{1}\IEEEauthorrefmark{2}, 
Robert A. Bridges\IEEEauthorrefmark{1}, 
Frank L. Combs\IEEEauthorrefmark{1}, 
\& Michael R. Moore\IEEEauthorrefmark{1}}
\IEEEauthorblockA{\IEEEauthorrefmark{1}Oak Ridge National Laboratory, Oak Ridge, TN,   
\IEEEauthorrefmark{2}Florida Atlantic University, Boca Raton, FL\\
ztyree@fau.edu, bridgesra@ornl.gov, combsfl@ornl.gov, mooremr@ornl.gov
 }
 }


%

%% file: 05-abstract.tex
\begin{abstract}
Modern vehicles rely on scores of electronic control units (ECUs) broadcasting messages over a few controller area networks (CANs). 
Bereft of security features, in-vehicle CANs are exposed to cyber manipulation and multiple researches have proved viable, life-threatening cyber attacks. 
Complicating the issue, CAN messages lack a common mapping of functions to commands, so packets are observable but not easily decipherable. 
We present a transformational approach to CAN IDS that exploits the geometric properties of CAN data to inform two novel detectors\textemdash one based on distance from a learned, lower dimensional manifold and the other on discontinuities of the manifold over time. 
Proof-of-concept tests are presented by implementing a potential attack approach on a driving vehicle. 
The initial results suggest that (1) the first detector requires additional refinement but does hold promise; (2) the second detector gives a clear, strong indicator of the attack; and (3) the algorithms keep pace with high-speed CAN messages.  
As our approach is data-driven it provides a vehicle-agnostic IDS that eliminates the need to reverse engineer CAN messages and can be ported to an after-market plugin.  
\end{abstract}

%% file: 10-intro.tex
\section{Introduction \& Background}
\label{sec:intro}

Modern vehicles 
rely on scores of embedded computers called electronic control units (ECUs) that control and facilitate communication of sub-systems by broadcasting messages over a few controller area network (CAN) buses. 
Since ECUs control most vehicle functions, adversarial manipulation of CAN signals have potentially severe consequences. 
CAN protocol is bereft of basic security features, e.g., encryption and authentication, and therefore vehicle CANs are exposed to exploitation. 
The now-mandatory OBD-II port gives direct access to CANs.  In addition the proliferation of vehicle interfaces, e.g., V2X, cellular, and WiFi, increases the attack surface. 
Multiple researchers have demonstrated remote access to CANs \cite{checkoway2011comprehensive, miller2015remote, woo2015practical} causing life-threatening manipulations, most notably, the remote Jeep hack of Miller \& Valasek \cite{miller2015remote}.

Consequently, recent research has focused on identifying CAN vulnerabilities and providing defensive capabilities. 
Many previous works have proposed changing CAN standards to incorporate security, e.g.,  \cite{szilagyi2010low, lin2012cyber, groza2013efficient}, but most are unadopted as they require costly evolution of standards and hardware.  
A notable exception, Brown et al.~\cite{brown2018can}, proposed message control firmware implemented in each ECU, a solution presumably cheap to retrofit. 
We pursue after-market,  vehicle-agnostic security solutions that do not require hardware changes.  

Developing vehicle-agnostic solutions is especially challenging for passenger vehicles because there is no publicly available translation from in-vehicle network data to vehicle functions. 
Further, CAN-to-function mappings vary per model. 
In short, even though packets are observable, there is no existing way to automatically know what mechanisms they control. 
This is a primary roadblock to defending CANs and calls for data analytics approaches to pioneer vehicle-agnostic detection and prevention capabilities \cite{jaynes2016automating, moore2017modeling, moore2018data, miller2013adventures}.

\subsection{Related CAN IDS Work}
Of the research to build intrusion detection systems (IDSs) for the currently-adopted CAN systems, there is a natural trichotomy: 
\begin{enumerate}[leftmargin = *]
\item \textbf{Rule-based IDS:} Early works involve rule-based detectors, an analogue of signature-based detection in enterprise security. These can detect simple signal-injection attacks, (sophisticated but very limited) bus-off attacks (see Cho et al.~\cite{cho2016error}), and potentially ECU reprogramming \cite{muter2010structured, hoppe2008security, hoppe2009applying}. 
\item \textbf{Frequency-based IDS:} Important CAN messages are sent redundantly with fixed frequency. 
The next trend in the CAN IDS research literature is to exploit the regular frequency of important CAN messages. Frequency anomalies have been explored to detect and prevent signal-injection attacks and potentially ECU reprogramming ~\cite{moore2017modeling, gmiden2016intrusion, song2016intrusion}. 
\item \textbf{ECU-fingerprinting IDS:} After publication of the Jeep hack \cite{miller2015remote}, in which some ECUs were silenced and others controlled to send spoofed messages in lieu of the silenced ECUs, CAN IDS research transitioned to more sophisticated methods all focused on automatic ECU identification as a stepping stone to IDS. (No sender/receiver information is included in CAN packets). These detectors use data-driven techniques to classify which ECU sent each message by exploiting timing or voltage signatures ~\cite{cho2016fingerprinting, leeotids, choi2018identifying}. These can detect signals that originate from the wrong transmitter. 
\end{enumerate}

Miller \& Valasek \cite{miller2013adventures, miller2015remote} exhibited the capability to compromise and control an ECU. 
It follows that Stuxnet-style attacks (see \cite{langner2011stuxnet}) on the CANs are possible and the logical next step for sophisticated adversaries. 
Specifically, ECU-authentic attacks, in which manipulated messages are sent from the expected ECU at the expected time, are possible. 
Such capabilities are exhibited by after-market ``chipping'', performance-tuning kits that reprogram ECUs. 
Albeit currently unseen, such attacks are not detectable by any above IDS, although possibly by 
one IDS that does not fit into the above trichotomy; Ganesan et al. \cite{2017-01-1654} use pairwise correlation of fixed values (e.g., speed, pedal angle) from CAN data and extra sensors to encode simple relationships and detect injection attacks.   
We build on this idea\textemdash using more sophisticated techniques to model subtle, nonlinear relationships in CAN data without needing to know it's functional meaning\textemdash with the goal of eventually proving these hypothetical attacks will be accurately detected.

\subsection{Novel Approach \& Contributions} 
To identify currently undetectable CAN manipulations, we present a transformational approach to CAN IDS.  
Our approach uses data-driven techniques to first learn, then exploit the shape\textemdash geometric and topological properties\textemdash of the CAN data to identify attacks. 
A vehicle's signals share necessary correlations because they communicate properties of a physical system. 
For a simple example, speedometer angle is highly correlated with the wheels' speed, but negatively correlated with braking. 
As modern vehicles comprise many complex, interacting systems, we expect more subtle and complicated relationships. 
Our hypotheses are that the naturally high-dimensional, time-varying CAN data will lie on a learnable, low dimensional manifold; 
the data will move continuously on this manifold over time; and even ECU-authentic, coordinated manipulations to CAN signals will cause detectable anomalies to these characterizations: 
\begin{enumerate}[leftmargin = *]
\item	\textbf{Distance from Manifold Anomaly:} Uncoordinated attacks (where malicious messages are not in concert with the car's other messages, such as the famous Jeep hack \cite{miller2015remote}) will produce data that leaves the manifold. 
\item 	\textbf{Time Increment Discontinuity:} Coordinated attacks (where all interdependent signals are spoofed to a valid, but manipulated configuration, as in a replay attack) and uncoordinated attacks will cause a discontinuity on the manifold if they change the vehicle's state drastically. 
\end{enumerate}

In this paper we introduce our novel, non-linear,  shape-based anomaly detection algorithms, provide visualizations of the time-varying CAN data manifold, and present initial results seeking proof-of-concept of this anomaly IDS by implementing a potential attack approach on a driving vehicle.   
Our algorithmic developments regard performance and are sufficiently advanced to admit online detection of messages on the high-speed CAN. 
Overall, this paper gives a transformational approach to CAN detection that has two main advantages: 1) it promises to detect ECU-authentic attacks that no published IDS  method has demonstrated; 2) as the approach is completely data-driven, it provides an after-market, vehicle-agnostic, and online IDS.

\subsection{CAN Message Basics} 
\begin{figure}[h]
\vspace{-.2cm}
\includegraphics[width=.48\textwidth]{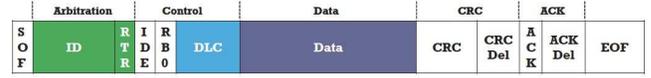}
\caption{
Image from Cho \& Shin \cite{cho2016fingerprinting} of CAN 2.0 data frame. 
11 bit arbitration ID field used for prioritization of messages. 
Up-to eight byte data field used for message contents.
}
\label{fig:CANframe}
\vspace{-.2cm}
\end{figure}

CAN is a broadcast protocol with 0 the dominant bit, meaning if multiple ECUs send opposing bits, the 0 overwrites the 1.  
All ECUs can emit bits until the end of the arbitration ID (AID) portion of the frame, whence the lone ECU with the lowest AID has the floor. 
(Each AID is assigned to a single ECU.)  
Refer to Fig~\ref{fig:CANframe}. 
The up-to eight byte data portion is the message contents, and injecting a signal requires only manipulating or injecting appropriate AID and data combinations to elicit the desired vehicle response. 
Most AIDs are sent with fixed frequency and redundant data. 

We follow a key discovery of Moore et al.~\cite{moore2018data}\textemdash \textit{for most signals the correct unit of analysis of CAN data are consecutive byte pairs of a data frame. } 
Each up-to eight-byte data frame padding with 0s if under eight bytes, decomposed into four separate two-byte messages with integer values in $[0, 2^{2*8}-1]$.  
I.e., most processes of a vehicle are indexed by the (AID, byte pair) combination; e.g., we have observed a single AID where the first byte pair encoded headlights while the last byte pair encoded speedometer values. 
We confirmed this claim both empirically and via consultation with a CAN engineer.  
Any CAN signals not following this convention will not be correctly encoded by our method. 
The mapping of byte pairs to function is non-public and varies per model; hence, our approach is data-driven.

%% file: 20-methodology.tex
\section{Methodology}
\label{sec:method} 
A priori, CAN traffic is high dimensional\textemdash vehicles often exhibit over 100 AIDs (over 400 byte pairs or PIDs), and each  taking potentially 
\begin{wrapfigure}[20]{r}{.26\textwidth}
\vspace{-.25cm}
\includegraphics[width=.28\textwidth]{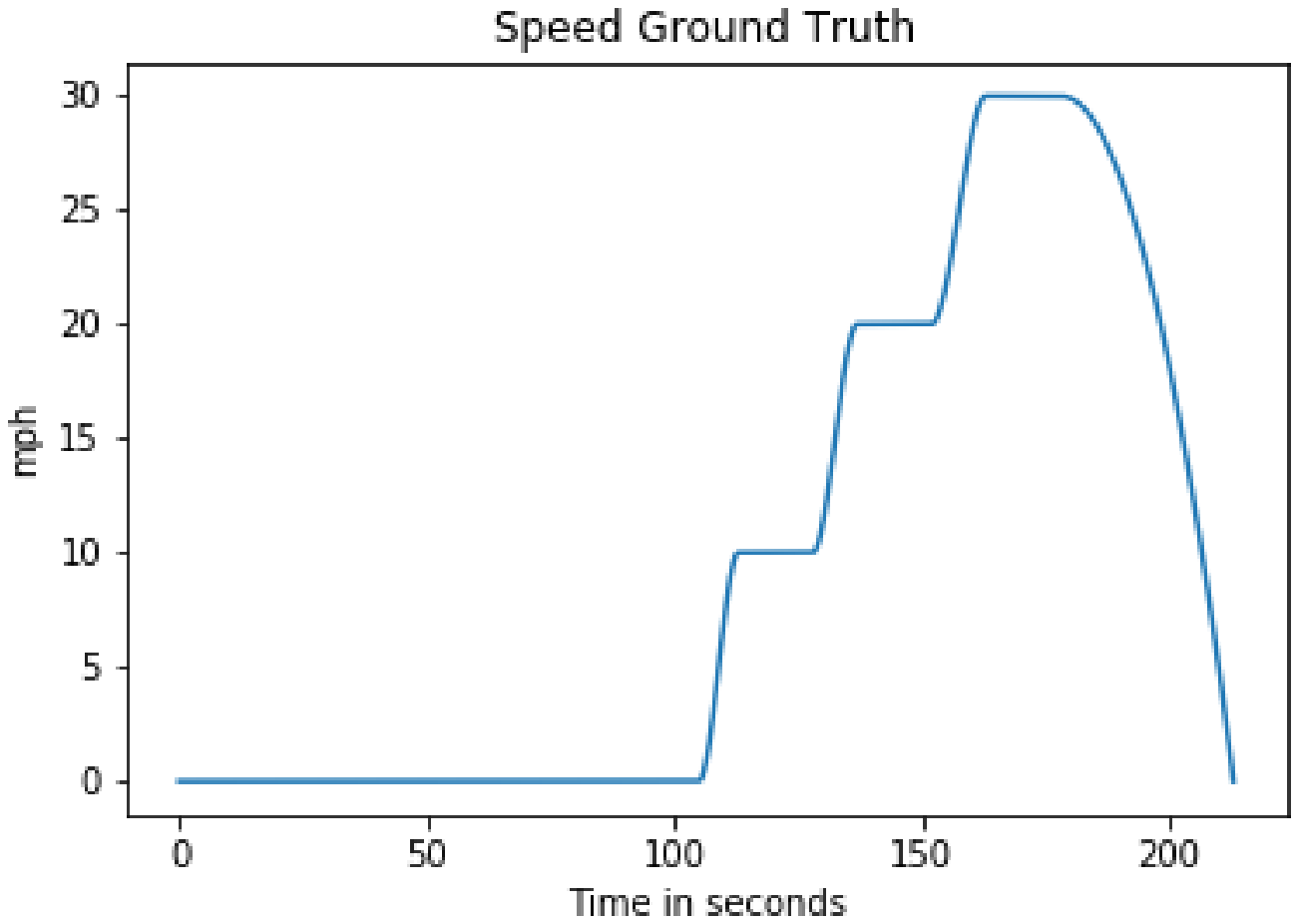}
\includegraphics[width=.28\textwidth]{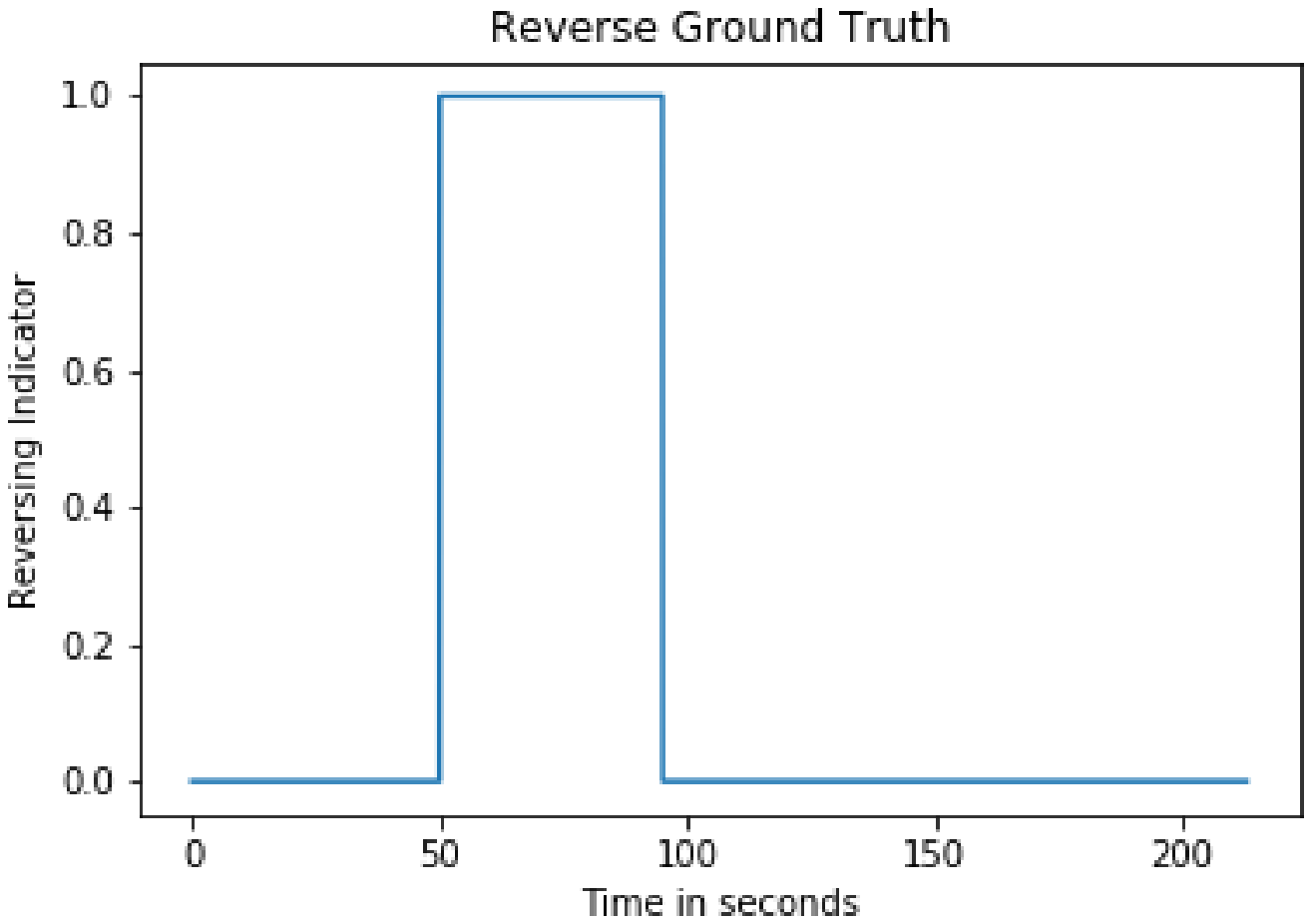}
\vspace{-.2cm}
\caption{Forward (top) \& Reverse (bottom) speed profiles for ambient data capture depicted.}
\label{fig:ground-truth-speed} 
\end{wrapfigure}
$2^{16}=65,536$ values. 
However, our understanding of the physical world suggests that there are far fewer actual degrees of freedom present in the data. 
This section presents a methodology to learn a more fitting, lower dimensional representation. 

As our application is anomaly detection, we require an ambient data capture for model fitting. Expecting different vehicle states to admit different inter-dependencies in the data, we train models on  data captured while (attempting to accurately) drive the vehicle in forward/reverse through the top/bottom speed profile in Fig. \ref{fig:ground-truth-speed}, respectively. 
Hence, the states we consider are Key On (on and in park), Accelerating, Speed (holding constant speed), Braking, and Reverse. 

For pre-processing, we discard any byte pairs for which only a single value is observed in all ambient data, leaving $\approx 70$ byte pairs.  
While our goal is initial implementation and proof of concept, actual use of our IDS will likely require greater variety in ambient data to minimize unseen signals. 
Each byte pair's values are regarded as a numerical time series.  

\input{22-clustering.tex}

\input{24-diffusion-manifold.tex}
\input{26-nystroem.tex}

\input{28-detectors.tex}

%% file: 22-clustering.tex
\subsection{Correlation Spectral Co-Clustering}

We apply spectral co-clustering of Dhillon \cite{dhillon2001co}, which treats a given data matrix as an adjacency matrix of a bipartite graph (with nodes for each row on one side, and nodes for each column on the other)  and uses normalized cuts (see Shi \& Malik \cite{shi2000normalized}) to find clusters. 
The upshot is both the rows and columns can be grouped simultaneously. 
For our application, we interpolate each series to a fixed length and generate the pair-wise correlation matrix. 
Equipped with this correlation matrix, we use spectral co-clustering implemented in Scikit~\cite{pedregosa2011scikit} to group byte pairs via correlation. 
In our application to the byte pair correlation matrix, each byte pair appears as a node on each side of the bipartite graph with edges weighted by the correlation of byte pair pairs. 

\begin{wrapfigure}[15]{r}{.28\textwidth}
\vspace{-.2cm}
\includegraphics[width = .28\textwidth]{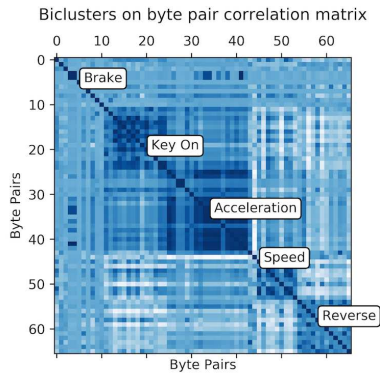}
\caption{Correlation heatmap after rearranging byte pairs to group co-clusters.} 
\label{fig:byte-pair-corr}
\vspace{-.2cm}
\end{wrapfigure} 


In order to label the clusters we include a synthetic ``canonical byte pair" for each state. 
E.g., for the ``Speed" state we take the time series of speed measurements during the data capture (see Moore et al.~\cite{moore2018data}), using a cubic spline, interpolate to the required byte pair length, and  cluster along with the actual byte pairs.
 
See Fig. \ref{fig:byte-pair-corr} produced by rearranging rows/columns to group clusters and apply vehicle state labels. 
The large darker regions along the diagonal represent groups of byte pairs with similar correlation structure. The clusters are labeled according to which cluster the above ``cannonical processes" belong. 
The main take-away is two-fold: (1) byte pairs comprising a cluster share a similar correlation structure across vehicle states; (2) There exists a mapping between clusters generated in this way and characteristics of the underlying physical system.

%% file: 24-diffusion-manifold.tex
\subsection{Learning a time-varying lower dimensional manifold}
Since clustered byte pairs vary together, we seek a low dimensional manifold for each cluster. 
This makes intuitive sense as we do not expect those byte pairs galvanized by accelerating to vary smoothly with those byte pairs stimulated upon braking and so on. 
For the sake of succinctness, we proceed only with the manifold for the ``Speed''  cluster. 
Let $p_1, ..., p_n$  denote the byte pairs of our cluster, and $x(t) := [ p_1(t), ..., p_n(t)]^T$, the vector of byte pair values at time $t$. 

\begin{figure}[h]
\vspace{-.4cm}
\includegraphics[width=.5\textwidth]{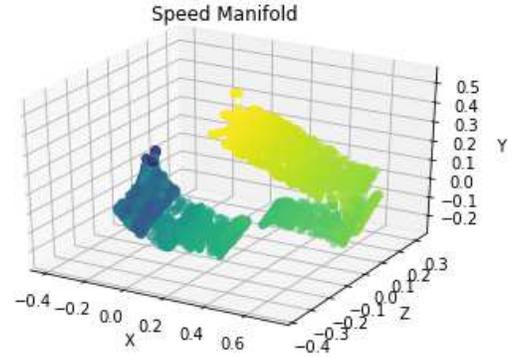}
\vspace{-.2cm}
\caption{Embedding of the ``Speed'' manifold into $m = 3$ dimensions. Colors yellow to green to blue indicate progression of the manifold over time. }
\label{fig:speed-manifold} 
\end{figure}

We use Diffusion Maps of Coifman \& Lafon \cite{coifman2006diffusion} to obtain a low dimensional representation for each time $t$ as follows:\footnote{See De la Porte et al.'s treatment \cite{de2008introduction} for an intuitive explanation.}  
\begin{enumerate}[leftmargin = *]
\item \textbf{Markov Marix} Let $P$ be the matrix so that $P(i,j)$, interpreted as the probability of jumping from $x_i$ to $x_j$, is proportional to $K(x_i, x_j) = \exp(-\gamma \|x_i-x_j\|^2).$ Here $\|. \|$ denotes Euclidean distance, and $\gamma$ a bandwidth parameter.\footnote{
Choosing $\gamma$ very small/large causes all points in the resulting embedding to be near/far from each other in the diffusion space. 
An effective heuristic we employed for choosing $\gamma$ is to plot $\log\sum P$ against $\log\gamma$ and choose $\gamma$ corresponding to the extent of the linear region. } 
\item \textbf{Diffusion Space} Map each $x_i \mapsto y_i: = [P^k(x_i *)]^T$, the $i^{th}$ column of $P^k$. 
Note that $P^k$ is the matrix of probabilities after $k$ hops, and $\sum_u (\|P^k(x_i,u) - P^k(u,x_j)\|^2) = \|y_i - y_j\|^2  $ is small for large $k$ iff $x_i$ and $x_j$ are close on the dataset's intrinsic geometry. 
\item \textbf{Projection} It follows that a lower dimensional representation respecting the data's shape is given by projecting to $m << n$ principle components of $P^k$ for large $k$ (equivalently of $P$, eliminating need for parameter $k$), where $m$ is chosen by the user.  
Explicitly, if $\xi_i$ are the $n$ eigenvectors of $P$ ordered by largest to smallest eigenvalues, then our projection is $x_i \mapsto \Psi(x_i): = \sum_{i=2}^{m+1} \langle y_i, \xi_i\rangle \xi_i.$ 
The first eigenvector is omitted as it is constant when $P = P^T$.
\end{enumerate}
Each color in Fig.~\ref{fig:speed-manifold} is the manifold obtained from the ``Speed" cluster for a single time $t$. 

%% file: 26-nystroem.tex
Naively, to embed a previously unseen point (e.g., $x(t+1)$), it is necessary to recalculate the Markov matrix and solve the eigenvector problem over again. 
While this poses no theoretical problems,
the matrix for to the ``Speed" cluster is on the order of $10^5 \times 10^5$, so recalculating the matrix at each time step to embed new points is infeasible for an online procedure.  
Leveraging our hypothesis, that the rank of the matrix is much lower than its size, we use the Nystr\"{o}m  method  \cite{williams2001using, yang2012nystrom}.

Concisely, we sub-sample $\hat x_1, \dots, \hat x_k$ for  $k = 1000 << 10^5 \approx n$  observed points, $\{x_i\}_{i = 1}^n$, and use the principle components of Nystr\"{o}m approximated $\hat P$ (an $n\times n$ approximation of $P$ with rank at most $k$).\footnote{
We define $\hat P$ from $\hat K: = ABA^T$, an approximation of $K$, where $A = (K(x_i,\hat x_j))_{n\times k} $ and $B$ the psuedo-inverse of $K(\hat x_i, \hat x_j)_{k \times k}$.}  
We choose $k=1000$ because at $k=1000$ we can embed points fast enough to keep up with the speed of the  CAN messages, and it seems likely that the dimension of the physical system determining the dependencies of the ``Speed" cluster is far less than 1000.

This permits calculation of the manifold incrementally at CAN message pace. 
Time is depicted in Fig.~\ref{fig:speed-manifold} by the progression from warm to cool colors. Note that the manifold evolves rather continuously.

%% file: 28-detectors.tex
\subsection{Detectors}
Our method gives rise to two methods of anomaly detection. 
\begin{enumerate}[leftmargin = *]
\item \textbf{Distance to Manifold Anomaly:} Given a previously unseen observation $x$ and $S$ the ambient dataset,  if $\|\Psi(x)- \Psi(S)|| > K_{\text{Dist}}$ we flag on $x_t$. 
Here $\Psi^{(k)}(S)$ denotes the embedding image of the ambient training set and $K_{\text{Dist}}$ a threshold. 
Intuitively, this schema reflects the idea that if we have a sufficient training set then normal observations should be mapped onto the manifold given by $\Psi(S).$  Conversely, if a signal is manipulated its discordance with the normal data should produce a jump from the manifold. 

\item \textbf{Time Increment Discontinuity:} The second schema involves continuity on the manifold. Let $x(t_i)$, $x(t_{i+1})$ denote subsequent observations. 
We flag if $||\Psi(x(t_i)) - \Psi(x(t_{i+1}))|| > K_{\text{Cont}}$. 
Intuitively, this schema reflects the idea that if an attacker  can manipulate all correlated PIDs in concert (as in a replay attack), they will be detected, unless they change the data in a fashion such that the resulting rate of change reflects that of the ambient dataset. 
\end{enumerate}

%% file: 30-experiment.tex
\section{Experiment} 
\label{sec:experiment} 
Here we present the initial test of our detectors. 
To implement a potential attack approach, 
we identified a collection of byte pairs that share high correlations to the vehicle speed (and therefore to each other). 
Using the OBD-II port, we injected at high frequency each of these byte pairs with a perturbation of its expected value while driving (on a dynamometer) at constant speed. 
If the perturbation is sufficiently large, this attack can disable a vehicle causing it to coast to a stop. Here we perturbed the values enough to cause the vehicle systems to malfunction but not shut down. 
The attack data capture is hosted over a period of 70s. 
Malicious traffic is injected in three 10s intervals, roughly $[10, 20], [30, 40], [50, 60].$

\begin{figure}[ht]
\hspace{-.4cm}
\includegraphics[width=9cm,height=4.5cm]{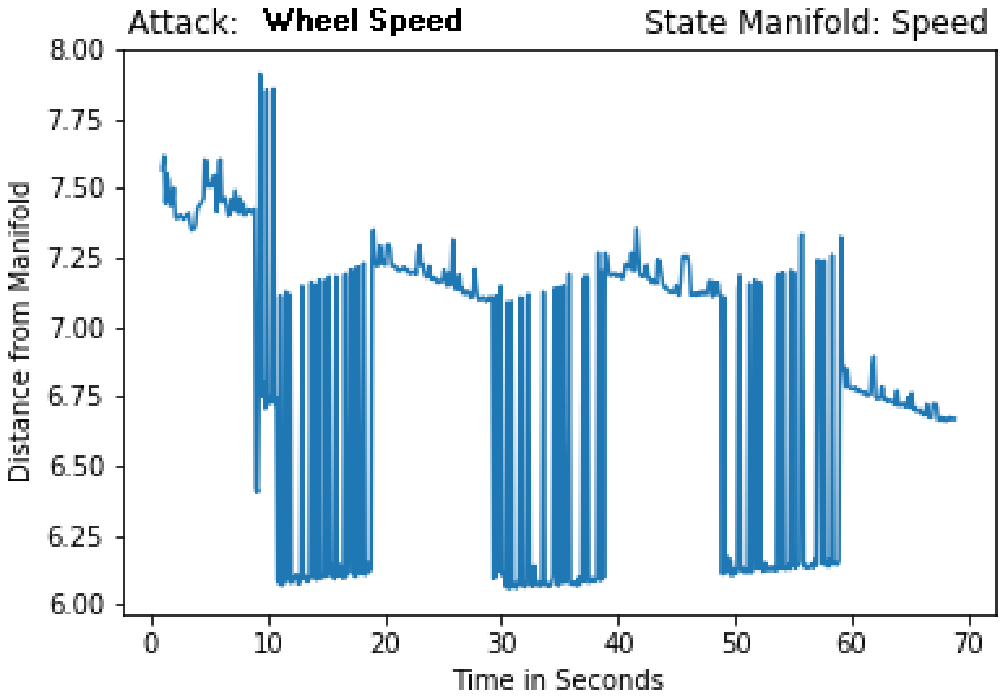}
\includegraphics[width=9cm,height=4.5cm]{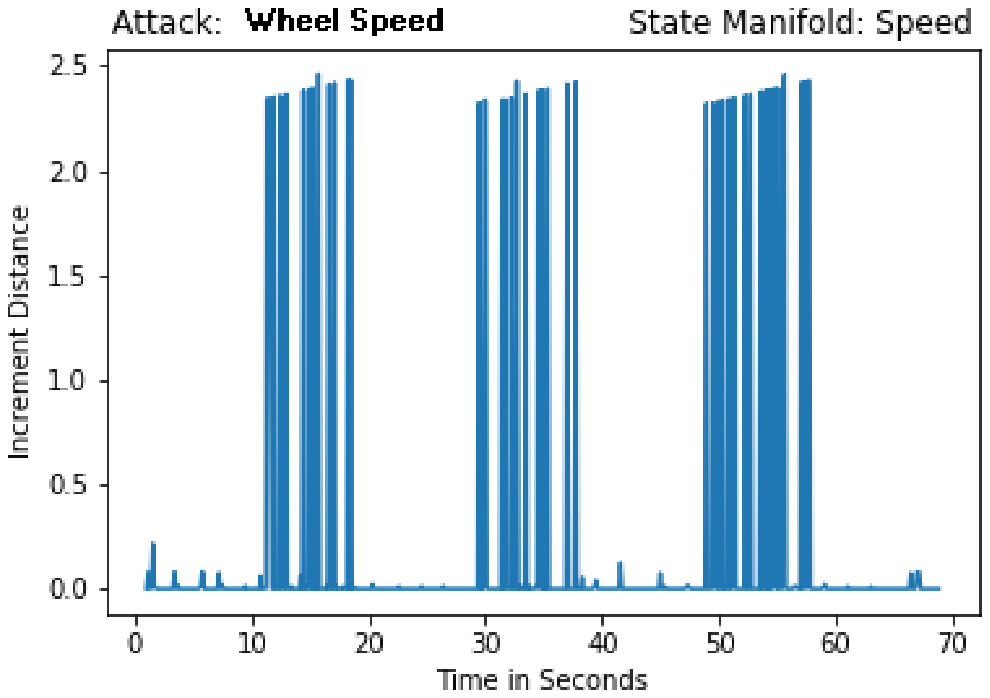}
\caption{Distance from ambient manifold (top) and increment distance (bottom) depicted for data with malicious values injected at high frequency for one byte pair during times $[10,20], [20,30], [30,40]$ seconds.}
\label{fig:results} 
\end{figure}

Fig.~\ref{fig:results} presents the two detection distances for the attack capture. The first shows the distance from the ``Speed" manifold of the CAN traffic over time. The seconds shows the distance between two consecutive points on the "Speed" manifold over time (referred to along the $y$-axis as increment distance). 
Both statistics were generated in an online fashion at faster or equal to the speed of CAN traffic.

The first figure exhibits unexpected behavior. 
While there is a clear signal during the attack period, the distance to the ambient manifold is often lower during attack intervals than otherwise. 
We conjecture that the ambient dataset is not sufficiently representative of the testing data capture.

The second figure, depicting increment distance, provides a clear signal of the attack, admitting accurate detection for a wide range of thresholds. 
In the event of a more sophisticated attack where ECU-authentic messages are sent with manipulated values, our evidence is that the oscillations seen here would not be present, and instead the figure would show six outlying plateaus each denoting the start or stop of an attack interval, respectively. 
Because this is an online procedure and messages on the CAN bus come milliseconds apart, this detection schema can identify an attack and alert almost instantly.
Both figures, particularly the second, suggest that the general shape-based approach does capture the information needed to for separating ambient and anomalous CAN traffic in a completely car agnostic fashion, but more refining and testing is needed.

%% file: 40-conclusion.tex
\section{Conclusion} 
\label{sec:conclusion} 
This paper introduces an online, manifold learning technique to produce two novel  anomaly detectors that seek to exploit unexpected changes in the geometric properties of CAN data for detection. 
From an ambient capture of CAN data, spectral co-clustering is used to group CAN data according to correlated signals during each vehicle driving state. 
Next, diffusion mappings are used to produce a low dimensional representation of the data for each state, and an update formula allows the manifolds to evolve over time. 
We discovered that this representation evolves continuously, and leverage this fact for detection. 
Our work informs two novel detectors: one based on distance from the expected manifold, learned in a training period on ambient data; the second based on increment distance in the lower dimensional space over time. 
Further the algorithm is crafted with sufficient treatment of computational complexity to permit real-time application. 
Targeting passenger vehicles, in which no publicly available mapping of the CAN signals to vehicle functions is present, our method is data driven, to facilitate after-market use upon sufficient refinement.

Seeking proof of concept, we implement a potential attack scenario on a driving vehicle, and test the detectors. 
In terms of accuracy, our results suggest that the first detector is currently insufficient and more representative training data is likely necessary. 
The second detector gives a strong indication of the attacks, working as expected. 
We note that this detector promises to identify ECU-authenticated attacks\textemdash where signals are sent from the expected ECU at the expected time, but with manipulated data. 
Performance-wise, the algorithm keeps pace with CAN data, meaning real-time detection is possible.

Further testing and refinement is needed. 
A deeper look at the manifold distance detector, in particular, investigations into what constitutes a sufficient training data set is necessary. 
For the increment distance, the results are overwhelmingly positive. 
Testing on more diverse and sophisticated attacks is needed, as well as on multiple diverse vehicles.  
For both detectors a method for setting the threshold is required. 
Overall, the paper introduces and provides proof-of-concept of a novel, computationally viable CAN IDS that holds the promise of detecting next generation CAN attacks.

%% file: 90-acknowledgements.tex
\section*{Acknowledgements}
\label{sec:acks} 
Authors' thank M. Verma and reviewers who helped refine this document. Research sponsored by the Laboratory Directed Research and Development Program of Oak Ridge National Laboratory, managed by UT-Battelle, LLC, for the U. S. Department of Energy and National Science Foundation Math Science Graduate Internship.